\documentclass[journal]{IEEEtran}
\pdfoutput=1
\IEEEoverridecommandlockouts
\usepackage{cite}
\usepackage{amsmath,amssymb,amsfonts}
\usepackage{algorithm,algorithmic}
\usepackage{bm}
\usepackage{mathrsfs}
\usepackage{multirow}
\usepackage{amsfonts,amssymb}
\usepackage{tabu}

\usepackage{array}

\usepackage{subfigure}
\usepackage{graphicx}
\usepackage{textcomp}
\usepackage{xcolor}
\def\BibTeX{{\rm B\kern-.05em{\sc i\kern-.025em b}\kern-.08em
    T\kern-.1667em\lower.7ex\hbox{E}\kern-.125emX}}

\usepackage{float}
\usepackage{hyperref}
\usepackage[hyphenbreaks]{breakurl}

\begin{document}


\title{STEP: Spatial-Temporal Network Security Event Prediction \\
}

\author{Qiumei Cheng,
	Yi Shen,
	Dezhang Kong,
        Chunming Wu,~\IEEEmembership{Member,~IEEE,}
\thanks{This work is supported by the National Key R$\&$D Program of China (2018YFB1800601), the Industrial Internet innovation and development project (TC190A449), the Key R$\&$D Program of Zhejiang Province (2020C01021), and the Major Scientific Project of Zhejiang Lab (2018FD0ZX01).}
\thanks{The authors are with the College of Computer Science and Technology, Zhejiang University, 310000, China (e-mail: chengqiumei@zju.edu.cn; shenyizju@zju.edu.cn; kdz@zju.edu.cn; wuchunming@zju.edu.cn). Chunming Wu is also with the Zhejiang Lab, 310000, China.}
}

\maketitle

\begin{abstract}
Network security events prediction helps network operators to take response strategies from a proactive perspective, and reduce the cost  caused by network attacks, which is of great significance for maintaining the security of the entire network. Most of the existing event prediction methods rely on temporal characteristics and are dedicated to exploring time series predictions, but ignoring the spatial relationship between hosts. This paper combines the temporal and spatial characteristics of security events and proposes a spatial-temporal event prediction model, named STEP. In particular, STEP formulates the security events prediction into a spatial-temporal sequence prediction. STEP utilizes graph convolution operation to capture the spatial characteristics of hosts in the network, and adopts the long short term memory (LSTM) to capture the dynamic temporal dependency of events. This paper verifies the proposed STEP scheme on two public data sets. The experimental results show that the prediction accuracy of security events under STEP is higher than that of benchmark models such as LSTM, ConvLSTM. Besides, STEP achieves high prediction accuracy when we predict events from different lengths of sequence.

\end{abstract}


\section{Introduction}
\IEEEPARstart{N}{etwork} security events are always generated by monitoring tools in the network (e.g., the intrusion detection systems), and are usually captured in the form of network logs for subsequent analysis by security analysts ~\cite{Liuyang2015}. Security events reflect the current security situation of the network system, which provides possibilities for security situation perception and attack intention inference. Meanwhile, it is possible to predict the upcoming security incidents of the system through data analysis of massive security events ~\cite{Liuyang2015, Musk2019}. 

Network security events prediction aims to predict possible intrusions from a proactive defense perspective. It enables the network defender to take proactive strategies instead of taking countermeasures after discovering an incident. which can facilitate network managers to make early warning systems before the system is compromised~\cite{Sunnan2019}. For example, in multi-step attacks, it is very important to predict the possible attack behavior before the attacker reaches the target.  At the same time, effective proactive prediction and defense can greatly reduce the potential cost of attack events. In this case, the prediction of security events is a supplement or augmentation to anomaly detection. In addition, the prediction of network security incidents can also help risk evaluation. Therefore, a robust prediction system will help improve detection and defense capabilities and also benefit the downstream task like intrusion response system.

Most of the existing methods learn attack behaviors from historical events, and these methods are based on the hypothesis that the event that occurs in the future will only be affected by the previous events. Considering that the attacker may change the corresponding behavior over time, researchers mostly use the obtained time characteristics to predict the events that may occur in the future\cite{Soldo2011,Shen2018,Park2012,zhenxin2015,Abdlhamed2016}. However, it is difficult to capture the spatial characteristics of the attacker based only on temporal dependencies. Specifically, the spatial characteristics here refer to the spatial correlation between IP addresses~\cite{chenyuzhong2015}.  For example, if an attacker keeps changing targets on the same day, and the infected hosts may be a stepping stone to compromise other devices. In this case, a simple temporal prediction model is unable to predict the following events. How to capture both temporal and spatial characteristics to predict security events in the network is still a huge challenge. 

\subsection{Literature Review}
Existing research uses different methods to infer potential future security events, and the methods vary from data types and data sources. A common method is to correlate events to find the causal relationship between events and use correlation rules to identify the context information of a security event. Given a particular event, it is possible to locate its attack scenario and reason about the next event~\cite{yonho2014}. If there is not enough historical data, probabilistic methods can also be used to make predictions. Wu \emph{et al.} ~\cite{Jinyu2012} propose a network attack prediction model with a Bayesian network, and they believe that environmental factors would affect the probability of an attack event. Therefore, the Bayesian network is applied to capture environmental factors and then make predictions.

In addition, researchers also pay attention to stochastic processes to predict security incidents. The authors in ~\cite{maochao2018} believe that the time interval of hacker intrusion events and the scale of intrusion should be modeled by random processes instead of data distribution. Specifically, they believe that  the interval between intrusion events can be described by a point process, and the evolution of the scale of intrusion can be described by AutoRegressive and Moving Average-Generalized AutoRegressive Conditional Heteroskedasticity (ARMA-GARCH). Their research shows that the stochastic process model can predict the interval and scale of event arrival. Similarly, Condon \emph{et al.} ~\cite{Condon2008} use Autoregressive Integrated Moving Average (ARIMA) to predict security events on a data set. Chen \emph{et al.} ~\cite{chenpeng2017} adopt a point process to predict the attack rate. 

Based on the observation of historical records, time series can also be utilized as one of the common methods for predicting security events. Park \emph{et al.} ~\cite{Park2012} propose a prediction mechanism, FORE, based on time series analysis and linear regression. FORE  predicts unknown Internet worms by analyzing the randomness of network traffic. The response speed of FORE to unknown worms is 1.8 times of ADUR. Zhan \emph{et al.} ~\cite{zhenxin2015} utilize long-term dependency features and extreme values to predict the number of attacks against honeypots. 

In recent years, the development of deep learning and data mining technology has improved the accuracy of predicting network intrusion behaviors. In  ~\cite{Soska2014}, Soska \emph{et al.} train a classifier on text data to predict whether the current normal webpage will become malicious in the short term. Liu \emph{et al.} ~\cite{Liuyang2015} explore the prediction problem of data leakage. They extract 258 features from more than 1,000 collected events, and achieve a true positive rate of 90$\%$ by using a random forest classifier. The characteristics studied include malicious DNS configuration, time series of malicious activities, and etc. Tiresias~\cite{Shen2018} uses LSTM to predict the actions an attacker will take in the future, especially from malicious events with noise. The research results show that recurrent neural networks (RNN) can accurately predict multi-step attacks and is better than other short-term memory-based models, such as Markov chains.

Xu \emph{et al.} ~\cite{xukuai2011} try to capture the similarity of the behavior of hosts with bipartite graph, and use a spectral clustering algorithm to divide hosts into clusters of different behaviors. The spatial correlation between hosts is also vital in security events prediction. Here, the spatial characteristic refers to the spatial correlation between IP addresses. Researchers have combined spatial-temporal characteristics for important areas such as network attack detection~\cite{Yepeng2019,wangwei2017} and prediction~\cite{chenyuzhong2015}. Inspired by the recommendation systems, Soldo \emph{et al.} \cite{Soldo2011}  predict the blacklist by combining temporal and spatial characteristics from historical logs. They used time series to capture the temporal characteristics in historical logs and capture spatial characteristics from the similarity of infected hosts and similarity between infected hosts and attackers. In ~\cite{chenyuzhong2015}, the authors study the time-related attack frequency within the continuous IP address and discover the inherent spatiotemporal pattern behind the cyber attack. They treat the IP address as a spatial variable and represent the attack frequency time series based on information entropy.

However, the existing spatial-temporal prediction schemes are unable to capture the characteristics of the neighboring nodes about a given host, and the host will be spatially affected by the surrounding hosts.

\subsection{Proposed Approach}
To address the above-mentioned challenges, we propose a spatial-temporal event prediction model, named STEP. In particular, STEP formulates the security events prediction into a spatial-temporal sequence prediction. STEP utilizes graph convolution operation to capture the spatial characteristics of hosts in the network and adopts the long short term memory (LSTM) to capture the dynamic temporal dependency of events. First, STEP establishes an undirected graph based on the interaction between hosts according to the history of network events. Each node in the graph is a host, and the edges of the graph represent the existence of communication records between hosts. Secondly, the STEP model introduces a graph convolutional network to aggregate the node information around the central node to form a new node representation. Network security events within a period can be regarded as a sequence. Therefore, this chapter models event forecasting as a time series forecasting problem. STEP performs a single-step prediction, which is to predict the event that will occur in the network at the next moment.
 
In summary, the contributions of this paper are manifolds:
\begin{itemize}
	\item We propose a novel network security event prediction (STEP) model from a spatial-temporal perspective. STEP is able to predict events that will occur in the network by capturing the temporal and spatial characteristics.
    \item We model event prediction as a time series forecasting problem and predict the next event based on the sequence within a given period of time. STEP allows for the single-step prediction based on historical data at any time.
    \item We have validated the proposed STEP scheme on two existing publicly available data sets. The experimental results show that the prediction accuracy of events under STEP is higher than that of benchmark models such as LSTM. Besides, STEP achieves high prediction accuracy when we predict events from different lengths of sequence.
\end{itemize}
The remainder of this paper is organized as follows. Section~\ref{overview} outlines the solution overview of STEP. Section~\ref{methodology} elaborates the methodology of our approach. In Section~\ref{experiment}, we evaluate the performance of the proposed approach with two benchmark data sets. Finally, Section~\ref{conclusion} concludes this paper and discusses future work.


\section{Solution Overview}
\label{overview}
\begin{figure*}[htb]
    \centering
    \includegraphics[width=0.8\linewidth]{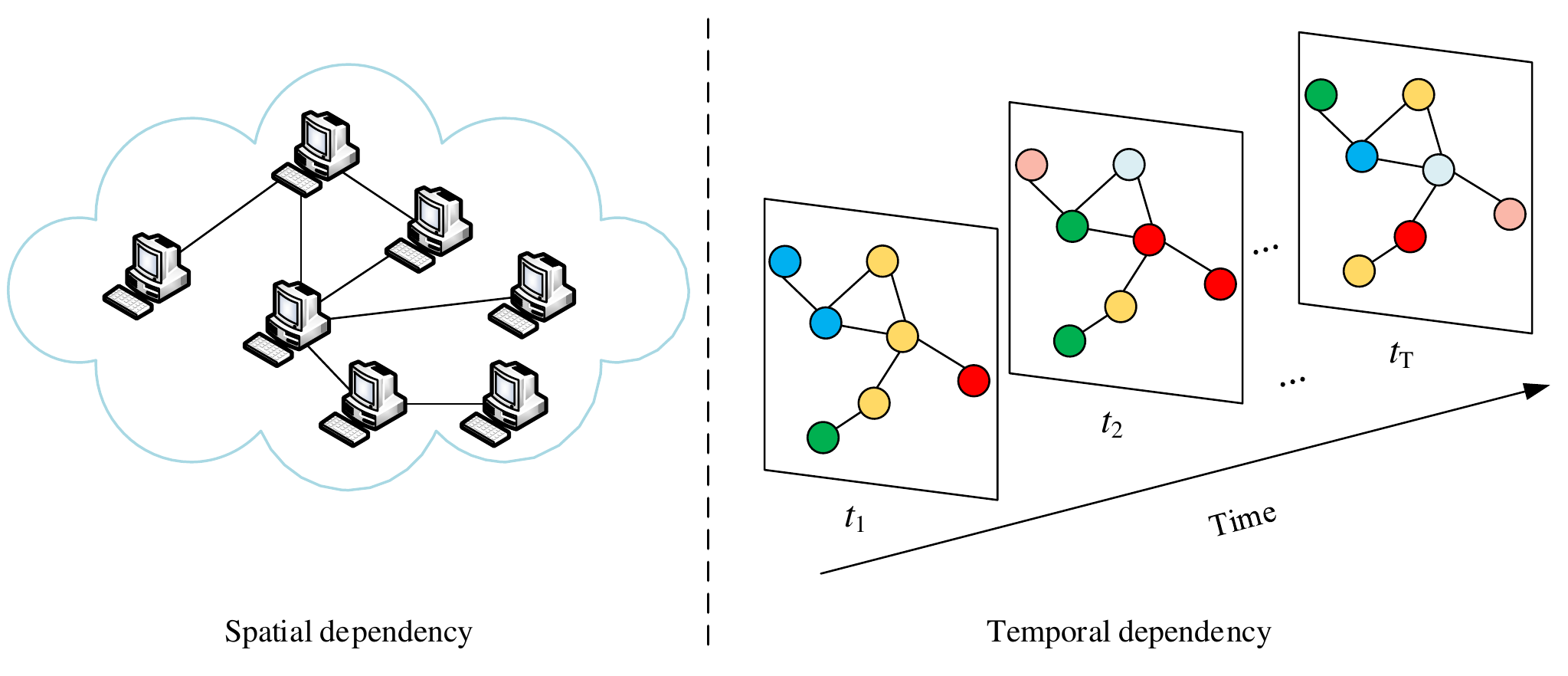}
    \caption{Framework of STEP}
    \label{ch4:fig:overview}
\end{figure*}

Fig.~\ref{ch4:fig:overview} demonstrates the temporal and spatial dependency of network security event prediction in STEP. As shown in Fig.~\ref{ch4:fig:overview}, we construct the undirected graph according to the history log records, where the spatial dependency means the correlation between hosts. The spatial correlation between host IP addresses has been used as a spatial feature and has been applied in attack prediction~\cite{chenyuzhong2015}. The right side of the figure shows the temporal dependency of security events. 

The events generated at each time step $t$ can be represented by an undirected graph, and the different colors of the nodes in the graph denote corresponding events on the node (host). The events that occur on the node are constantly changing over time. Specifically, the events generated by the host in the graph at each time step $t$ can be regarded as a sequence. Thus, given a period of time $s$, the events prediction on the host at the next time step is essentially a time series prediction problem. In particular, STEP uses LSTM~\cite{hochreiter1997long} to capture time dependency, and performs graph convolution operation to capture the spatial characteristics between hosts. The STEP model contains two parts of input: a sequence that changes over time; an undirected graph containing the spatial relationship between hosts.  The constructed undirected graph can be considered as the prior knowledge in the event prediction.

First, STEP defines the network as an undirected graph $\mathcal{G}=\left ( \mathcal{V},\mathcal{E} \right )$, where $\mathcal{V}$ is the set of the hosts in the network, and $\left | \mathcal{V} \right |=N$ is the total number of hosts, and $\mathcal{E}$  represents the set of the edges on the graph. The edges on the graph are created based on the interactions between hosts in the network history log. At each time step $t$, the events generated by the hosts on the graph can be represented by $\mathbf{X}_{t}=\left ( \boldsymbol{x}_{t}^{1}, \boldsymbol{x}_{t}^{2},\cdots ,\boldsymbol{x}_{t}^{i}, \cdots ,\boldsymbol{x}_{t}^{N},\right )^{T}\in \mathbb{R}^{N \times d}$, where $\boldsymbol{x}_{t}^{i}\in \mathbb{R}^{d}$ denotes the event of the node $i$ at the time step $t$, and $d$ is the feature dimension of the event.

According to above security events formulation, the prediction task of the security events of the entire network can be defined as follows. Given a period of time $T$,  the historical events data can be represented by $\mathcal{X}_{T}=\left ( \mathbf{X}_{1},\mathbf{X}_{2},\cdots ,\mathbf{X}_{T} \right )\in \mathbb{R}^{N \times d \times T}$. If we choose the events data in the time slice $s$, $\mathcal{X}_{s}=\left ( \mathbf{X}_{t-s+1},\mathbf{X}_{t-s+2},\cdots ,\mathbf{X}_{t} \right )\in \mathbb{R}^{N \times d \times s}$, we can predict the events that will occur in the next $p$ moments $\mathcal{Y}=\left ( \mathbf{X}_{t+1},\mathbf{X}_{t+2},\cdots ,\mathbf{X}_{t+p}\right )\in \mathbb{R}^{N \times d \times p}$. This paper mainly focus on the security events on the hosts at the next moment, so we only consider security event prediction in the case of $p=1$, i.e., single step prediction. At this time, $\mathcal{Y}=\ mathbf{X}_{t+1}\in \mathbb{R}^{N \times d }$.

\section{Methodology}
\label{methodology}
In this section, we detail the methodology of STEP for security event prediction. Specifically, we introduce theoretical knowledge of graph convolutional networks and long short-term memory. Then, we elaborate on the cell of STEP.

\subsection{Graph Convolutional Networks}
The core idea of the graph convolutional network (GCN) is to aggregate node information from stacked graph convolutional layers, thereby generating new node representations. The implementation of GCN can be divided into spatial domain and spectral domain. The spatial domain-based methods are mainly to find the neighbors of the central node, and then extract the features on the neighbors \cite{Niepert2016}. Recently, most popular GCN models use spectral domain knowledge to extract spatial features on topological graphs. The general idea behind the spectral method is to first transform the signal on the graph into the spectral domain, and then convert it back to achieve graph convolution after spectral domain processing. 

Generally, for a graph with $n$ nodes, each node has a specific value, and the signal of the graph can be regarded as a $n$-dimensional vector composed of the values on these $n$ nodes. Defferrard \emph{et al.} \cite{Defferrard2016} gave the spectrum definition of the graph convolution operation $*_{\mathcal{G}}$. Given an undirected graph $\mathcal{G}=\left (\mathcal{V},\mathcal{E},A \right )$, where $\left | \mathcal{V} \right |$ is the set of nodes, and $\left | \mathcal{E} \right |$ is the set of edges, and $A$ is the adjacency matrix. The signal $x\in \mathbb{R}^{n\times d_{x}}$ on the graph corresponds to the feature vector of the node on the graph, and $d_{x}$ is the dimension of the feature vector. 

Generally, the transformation of the graph signal in the spectral domain requires the Laplacian matrix of the graph. A standard Laplacian matrix can be defined as: $L=I_{n}-D^{-\frac{1}{2} }AD^{-\frac{1}{2}}\in \mathbb{R}^{n \times n}$, where $I_{n}$ is the identity matrix, and $D\in \mathbb{R }^{n \times n}$ is the degree matrix, with $D_{ii}=\sum_{j}A_{ij}$. The Laplacian matrix is a real symmetric semi-definite matrix. According to the properties of the Laplacian matrix, it can be known that its eigenvectors can be orthogonal to each other. Moreover, the Laplacian matrix can be spectrally decomposed, and its spectral decomposition form is: $L=U \Lambda U^{-1}$, where $U=\left [u_{0},u_{1},\cdots ,u_{n-1} \right] \in \mathbb{R}^{n\times n}$ is a matrix composed of eigenvectors, and $\Lambda=diag\left (\left [\lambda _{0}, \lambda _{1},\cdots ,\lambda _{n-1}\right] \right) \in \mathbb{R}^{n\times n}$ is the diagonal matrix that composed of the eigenvalues of the standardized Laplacian matrix. Because $L$ is an orthogonal matrix, its spectral decomposition can be written as $L=U \Lambda U^{T}$. When transforming the graph signal to the spectral domain, the set of bases required by the Fourier transform is the eigenvector of the Laplacian matrix. This is because the Laplacian matrix has $n$ linearly independent orthogonal vector.

Given a graph signal $x\in \mathbb{R}^{n\times d_{x}}$, its Fourier transform on the graph is:
\begin{equation}
\hat{x}=U^{T}x \in \mathbb{R}^{n}
\end{equation}
Further, its inverse Fourier transform is: $x=U\hat{x} \in \mathbb{R}^{n}$. According to the convolution theorem, the Fourier transform of a function convolution is the product of the function Fourier transform. Further, by extending the convolution theorem to the graph, the graph convolution of the signal can be calculated. The graph convolution of the two signals $x$ and $y$ can be expressed as:
\begin{equation}
x*_{\mathcal{G}}y=U\left ( \left ( U^{T}x \right )\odot \left ( U^{T}y \right ) \right )
\end{equation}
where $*_{\mathcal{G}}$ denotes the graph convolution, and $\odot$ is the Hadamard product. When the graph signal $x$ passes $g_{\theta }$, we are:
\begin{equation}
\label{equa:gcn1}
y=g_{\theta }*_{\mathcal{G}}x=g_{\theta }\left ( L \right )x=g_{\theta }\left ( U \Lambda U^{T} \right )x=Ug_{\theta }\left ( \Lambda \right )U^{T}x
\end{equation}
where $\theta \in \mathbb{R}^{n}$ is a vector of Fourier coefficients, and $g_{\theta }\left (\Lambda \right )=diag\left (\theta \right) $. However, Eq.~\ref{equa:gcn1} may induce a large computational cost. Especially for large graphs, the calculation cost of the eigendecomposition of the Laplacian matrix is also relatively high with $\mathcal{O}\left (n^{ 2} \right )$. To address the above issues, a common method is to approximate the convolution kernel $g_{\theta }\left (\Lambda \right )$ from the expansion of Chebyshev polynomials~\cite{Hammond2011 , Defferrard2016}. Using the $K-1$ truncated expansion of the Chebyshev polynomial to approximate the convolution kernel, we can get:
\begin{equation}
\label{equ:Lambda}
g_{\theta }\left ( \Lambda \right )= \sum_{k=0}^{K-1}\theta _{k}T_{k}\left ( \tilde{\Lambda} \right )
\end{equation}
where $T_{k}\left (\cdot \right )$ is the Chebyshev polynomial of order $k$, and $\theta \in \mathbb{R}^{K}$ is the Chebyshev coefficient vector, and $\tilde{\ Lambda}$ is the transformed diagonal matrix of eigenvalues: $\tilde{\Lambda}=2 \Lambda /\lambda _{max}-I_{n}$, and $\lambda _{max}$ is the maximum eigenvalue of Laplacian matrix. Since the Laplacian matrix is a semi-positive definite matrix and the eigenvalues are non-negative, the transformation of $\Lambda \geqslant 0$, $\tilde{\Lambda}$ is to ensure that the input of the Chebyshev polynomial is in $\left [-1 ,1 \right ]$.

Substitute the Eq.~\ref{equ:Lambda} into the Eq.~\ref{equa:gcn1}, we can get:
\begin{equation}
\label{eq:gcnCheby}
\begin{split}
y&=g_{\theta }*_{\mathcal{G}}x\\
&=U \sum_{k=0}^{K-1}\theta _{k}T_{k}\left ( \tilde{\Lambda} \right )U^{T}x\\
&=\sum_{k=0}^{K-1}\theta _{k}T_{k}\left ( U\tilde{\Lambda}U^{T} \right )x\\
&=\sum_{k=0}^{K-1}\theta _{k}T_{k}\left ( \tilde{L} \right )x
\end{split}
\end{equation}
where, $\tilde{L}=2L/\lambda _{max}-I_{n}$ and $T_{k}\left (\tilde{L} \right )$ is the $k$-order Chebyshev polynomial on the transformed Laplacian matrix $\tilde{L}$. In particular, $T_{k}\left (x \right )$ can be obtained by the recursive formula of Chebyshev polynomial. Generally, the recursive formula of Chebyshev polynomial can be defined as:
\begin{equation}
T_{k}\left ( \tilde{L} \right )=2\tilde{L}T_{k-1}\left (  \tilde{L}\right )-T_{k-2}\left ( \tilde{L} \right ) 
\end{equation}
\begin{equation}
T_{0}\left ( \tilde{L} \right )=I,T_{1}\left ( \tilde{L} \right )=\tilde{L}
\end{equation}
In this way, the computational complexity is $\mathcal{O}\left (K\left | \mathcal{E} \right | \right )$, which will be much smaller than that of Eq.~\ref{equa:gcn1}. $K$ refers to the number of steps from the center vertex, which can be seen as the receptive field of the convolution kernel, and each graph convolution will use $K$ steps around the center vertex, and the weight coefficient is $\theta_{k}$.

\subsection{LSTM}
Long short term memory (LSTM) is a popular variant of recurrent neural network (RNN) and has been widely used in recent years due to its advantages of long-term dependencies in learning sequence modeling tasks. In this paper, we adopt LSTM to predict the security events that will occur in the next moment. In particular, it is formulated into a sequence generation problem. A typical LSTM cell contains three kinds of gates, including the forget gate $f_{t}$, the input gate $i_{t}$ and the output gate $o_{t}$. Given an input $\mathbf{x}_{t}$, the forget gate  $f_ {t}$ determines which information of the previous unit state $ c_ {t-1} $ should be forgotten. Meanwhile,  the input gate $ i_ {t } $ controls the updated value. The hidden state $ h_ {t} $ is controlled by the output gate $ o_ {t} $. The introduction of a memory cell prevents the gradient from disappearing quickly. A typical LSTM cell can be defined as follows:
\begin{equation}
\begin{split}
f_{t}&=\sigma\left ( \mathbf{W} _{f}\cdot x_{t}+\mathbf{W} _{f}\cdot h_{t-1}+b_{f}  \right )\\
i_{t}&=\sigma\left ( \mathbf{W} _{i}\cdot x_{t}+\mathbf{W} _{i}\cdot h_{t-1}+b_{i}  \right )\\
\tilde{c} _{t}&=\tanh \left ( \mathbf{W} _{c}\cdot x_{t}+\mathbf{W} _{c}\cdot h_{t-1}+b_{c}  \right ) \\
c_{t}&=f_{t}\odot c_{t-1}+i_{t}\odot \tilde{c} _{t}\\
o_{t}&=\sigma\left ( \mathbf{W} _{o}\cdot x_{t}+\mathbf{W} _{o}\cdot h_{t-1}+b_{o}  \right )\\
h_{t}&=o_{t} \odot \tanh \left ( c_{t} \right )
\end{split}
\end{equation}
where $ \odot $ represents the Hadamard product, $ \sigma \left( \cdot \right)$ is the sigmoid activation function, $ b_ {f} $, $ b_ {i} $, $ b_ {c} $, $ b_ {o} $ is the biases. Given a period of historical events data, $\mathcal{X}_{s}=\left ( \mathbf{X}_{t-s+1},\mathbf{X}_{t-s+2},\cdots ,\mathbf{X}_{t} \right )\in \mathbb{R}^{N \times d \times s}$, LSTM is able to predict variable-length event sequence. In this way, LSTM allows for capturing temporal dependencies between different events.

\subsection{STEP cell}
In this paper, we propose a spatial-temporal approach to predict security events. Specifically, we learn from the spatial-temporal model from the literature ~\cite{Youngjoo2017}. In ~\cite{Youngjoo2017}, one solution is to stack graph convolutional neural networks and then use the output of GCN as the  input of LSTM to capture the dynamic temporal and spatial features of data. This paper utilizes another solution by generalizing the idea of convLSTM~\cite{ConvLSTM2015} to the graph. The core idea of the model is to replace the traditional Euclidean 2D convolution in convLSTM as the graph convolution $*_{\mathcal{G}}$.
\begin{equation}
\centering
\begin{split}
i_{t}&=\sigma \left ( \mathbf{W} _{xi}\ast _{\mathcal{G}}x_{t}+ \mathbf{W} _{hi}\ast _{\mathcal{G}}h_{t-1}+b_{i}\right )\\
f_{t}&=\sigma \left ( \mathbf{W} _{xf}\ast _{\mathcal{G}}x_{t}+ \mathbf{W} _{hf}\ast _{\mathcal{G}}h_{t-1}+b_{f}\right )\\
c_{t}&=f_{t}\odot c_{t-1}+i_{t}\odot \tanh\left ( \mathbf{W} _{xc}\ast _{\mathcal{G}}x_{t}+ \mathbf{W} _{hc}\ast _{\mathcal{G}}h_{t-1}+b_{c} \right )\\
o_{t}&=\sigma \left ( \mathbf{W} _{xo}\ast _{\mathcal{G}}x_{t}+ \mathbf{W} _{ho}\ast _{\mathcal{G}}h_{t-1}+b_{o}\right )\\
h_{t}&=o_{t}\odot \tanh\left ( c_{t} \right )
\end{split}
\end{equation}
where, $*_{\mathcal{G}}$ is the graph convolution operation shown in Eq.~\ref{eq:gcnCheby}, $\mathbf{W}_{h\cdot }\in \mathbb{R}^{K\times d_{h}\times d_{h}}$ is the weight coefficient, $K$ is the number of steps from the central node of the Chebyshev polynomial, $d_{h}$ is the corresponding hidden layer dimension. $\mathbf{W}_{x\cdot }\in \mathbb{R}^{K\times d_{h}\times d_{x}}$, where $d_{x}$ is the dimension of the vertex feature in the graph, i.e.,  the dimension of the event feature in this paper. 

The number of parameters is related to the $K$ in the Chebyshev polynomial, and has nothing to do with the number of vertices. For each step of graph convolution operation (e.g., $\mathbf{W} _{xi}\ast _{\mathcal{G}}x_{t}$), $K$ controls the cost of calculation, and each vertex needs to calculate the features of the neighbor of $K$-order. In this way, the central vertex can aggregate the information of its neighbor nodes. In this article, we set $K=2$, that is, the information characteristics of nodes whose center node is two steps away from each other. 

\begin{figure}[htb]
    \centering
    \includegraphics[width=\linewidth]{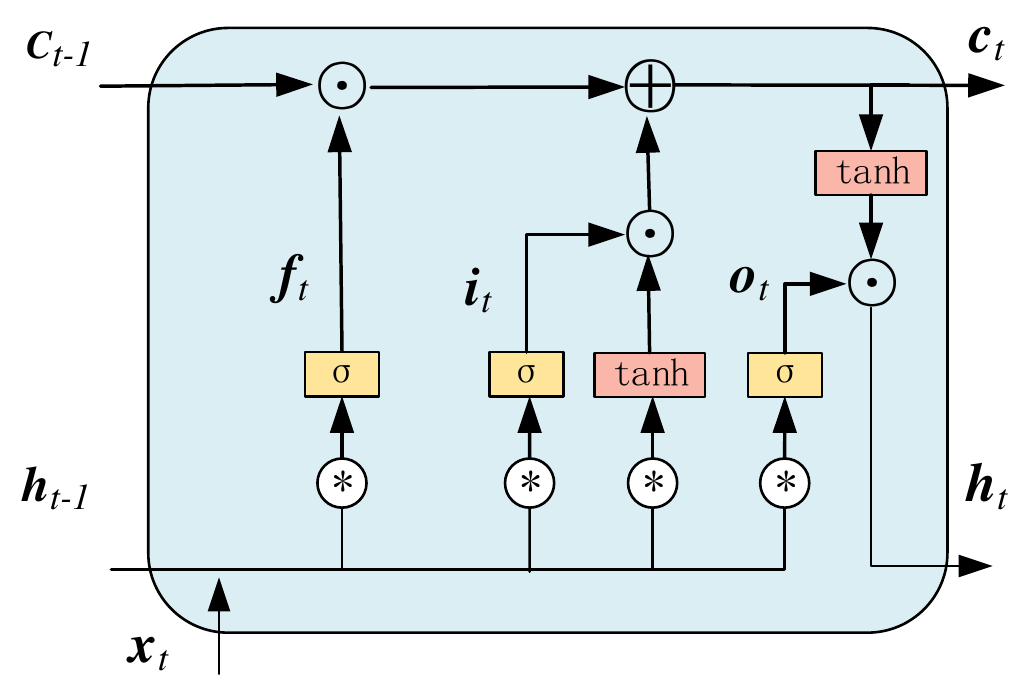}
    \caption{The STEP cell}
    \label{ch4:fig:gcnlstm}
\end{figure}

Fig.~\ref{ch4:fig:gcnlstm} illustrates the detailed STEP cell, which combines graph convolution operations in LSTM. The symbol $\circledast $ in the figure represents the graph convolution operation. The schematic diagram of the structure of the GCN-LSTM unit is shown in Figure \ref{ch4:fig:gcnlstm}. The symbol $\circledast $ in the figure represents the graph convolution operation. The main difference from the traditional LSTM unit is that each calculation involves a graph convolution operation.

\section{Evaluation}
\label{experiment}
In this section, we evaluate the performance of STEP in predicting security events. In particular, we select two famous event data sets LANL and HDFS to evaluate the performance of the proposed model. Besides, we also compare the performance with other baseline models.

\subsection{Dataset}
\textbf{LANL}~\cite{LANLdataset}: Los Alamos National Laboratory (LANL) collected the network and host events of the enterprise network within 90 days. The host event data set contains events that occur on all hosts running Windows operating systems on the LANL corporate network. This paper selects a subset of this dataset, containing 999,791 host events, about  9.21G. All the source event data is captured in JSON format. The event contains 21 fields, such as the timestamp, the ID of the event, and the name of the user who initiated the event. Table ~\ref{tab:LANL} shows the specific description of some events in the LANL data set.
\begin{table}[htb] 

\caption{Event description in LANL data set} 
\begin{center}
\label{tab:LANL} 
\begin{tabular}{p{1.5cm}|p{7cm}} 

\hline  
\hline  
EventID & Event Description \\ 
\hline 
1100&Event logging service has shut down\\  
\hline
4624&An account was successfully logged on\\
\hline
4625& An account failed to logon on\\
\hline
4648& A logon was attempted using explicit credentials\\
\hline
4672&Special privileges assigned to a new logon\\
\hline  
4768&Kerberos authentication ticket was requested (TGT)\\
\hline
4769&Kerberos service ticket was requested (TGS)\\
\hline
4770&Kerberos service ticket was renewed\\
\hline
4776&The domain controller attempted to validate the credentials for an account\\
\hline
4800&The workstation was locked\\
\hline
4801&The workstation was unlocked\\
\hline
4802 & The screensaver was invoked\\
\hline
4803&The screensaver was dismissed\\
\hline  
\hline  
\end{tabular}  
\end{center}
\end{table}

\textbf{HDFS}~\cite{HDFSdataset}: This data set is Hadoop Distributed File System (HDFS) event data in a private cloud environment. This data set does not contain any private information. We select 104,816 events, each of which contains 6 fields, such as Time, Pid, Level, Component, Content, EventID, and etc. Table ~\ref{tab:HDFS} demonstrates some key fields of the data.

\begin{table*}[htb] 

\caption{Examples of events data in HDFS data set} 
\begin{center}
\begin{tabular}{p{1.5cm}|p{1.5cm}|p{10cm}} 

\hline  
\hline  
Time & EventID & Event Content \\ 
\hline 
203531 & E5& Receiving block blk\_5647760196018207394 src: /10.251.127.243:46722 dest: /10.251.127.243:50010\\  
\hline  
203531 & E3& 10.251.71.193:50010 Served block blk\_-1608999687919862906 to /10.251.30.101\\  
\hline
203531 & E18 & 10.251.31.5:50010 Starting thread to transfer block blk\_-1608999687919862906 to 10.251.90.64:50010\\
\hline  
\hline  
\end{tabular}  
\label{tab:HDFS}  
\end{center}
\end{table*}

\subsection{Preprocessing}

The raw LANL data set is saved in JSON format, which contains a total of 21 fields. First, we converted it to CSV format and removed duplicates. The selected data set contains a total of 1423 time steps about events. Some hosts only generate events with extremely low probability in these 1423 time steps, which will lead to extremely sparse event prediction when creating the graph adjacency matrix. To address these sparse issues, we first count the frequency of occurrences of all hosts and then remove the hosts that a number of occurrences below 10. Finally, we obtain the event data of 261 hosts. Based on the processed host event data, we created the corresponding graph adjacency matrix.

\begin{figure}[htb]
    \centering
    \includegraphics[width=0.8\linewidth]{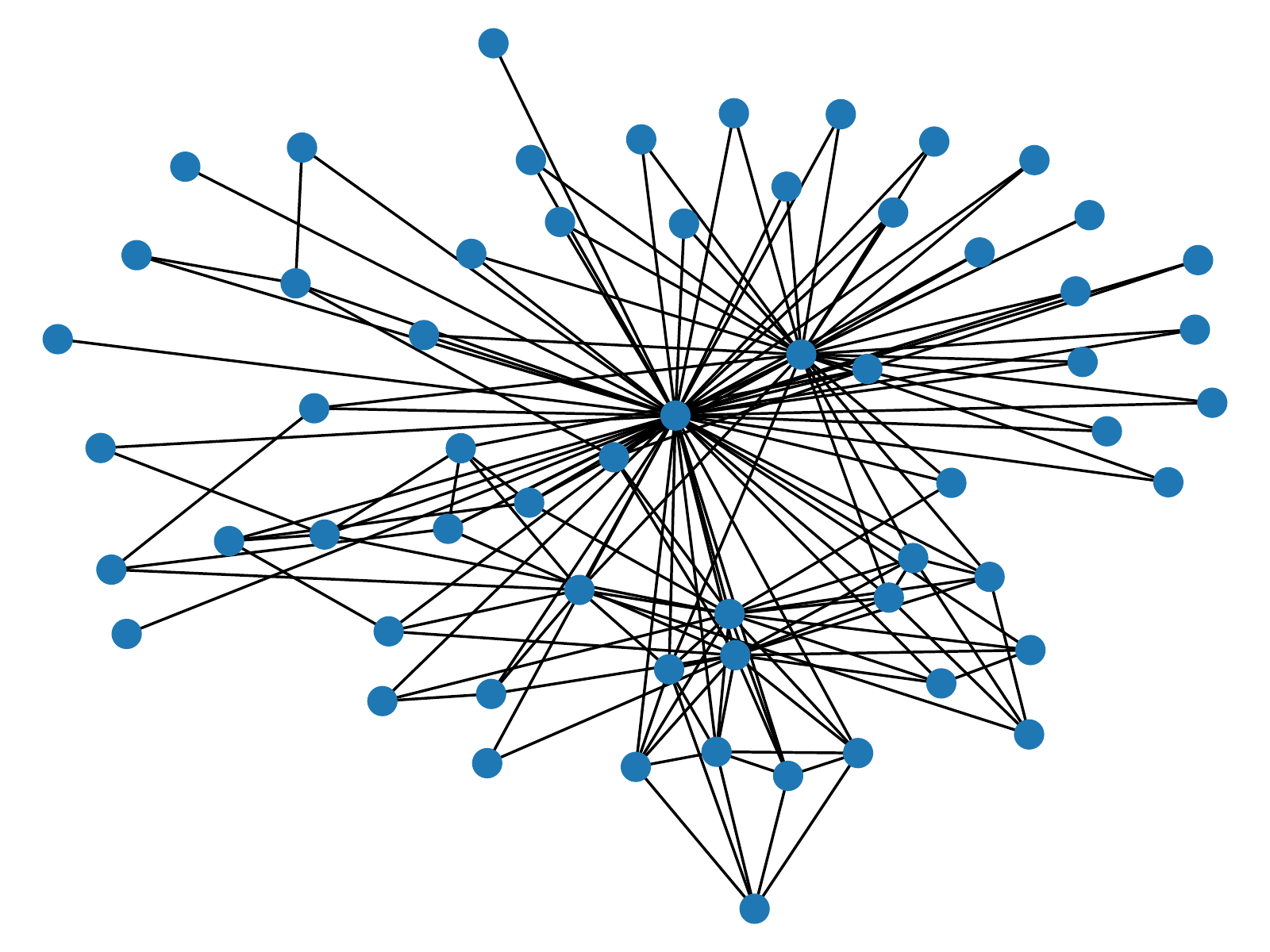}
    \caption{The undirected graph of partial nodes in LANL data set}
    \label{ch4:fig:lanl}
\end{figure}

Fig.~\ref{ch4:fig:lanl} shows the undirected graph of partial nodes in LANL. However, the events of the hosts in each time step are very sparse. For example, at time step $t$, only some hosts generate events, and none of the other hosts generate events. Specifically, we define the '0 event' that represents no event that occurs in the host. In this way, the feature matrix will be very sparse, and the events generated by most hosts at each moment are '0' rather than other event IDs.

In order to further reduce the sparsity of host events, we integrate the events in $k$ steps into a new time step. First, we find the event ID with the highest frequency in the $k$ step. If there is only one event with the highest frequency, then we use this event ID as the event on the new time step. If there are multiple highest frequencies that the event occurs, the most recent event is selected as the new event of the new time step. We finally obtained a total of 474 new time steps. The total number of events included is 86.

\begin{figure}[htb]
    \centering
    \includegraphics[width=0.8\linewidth]{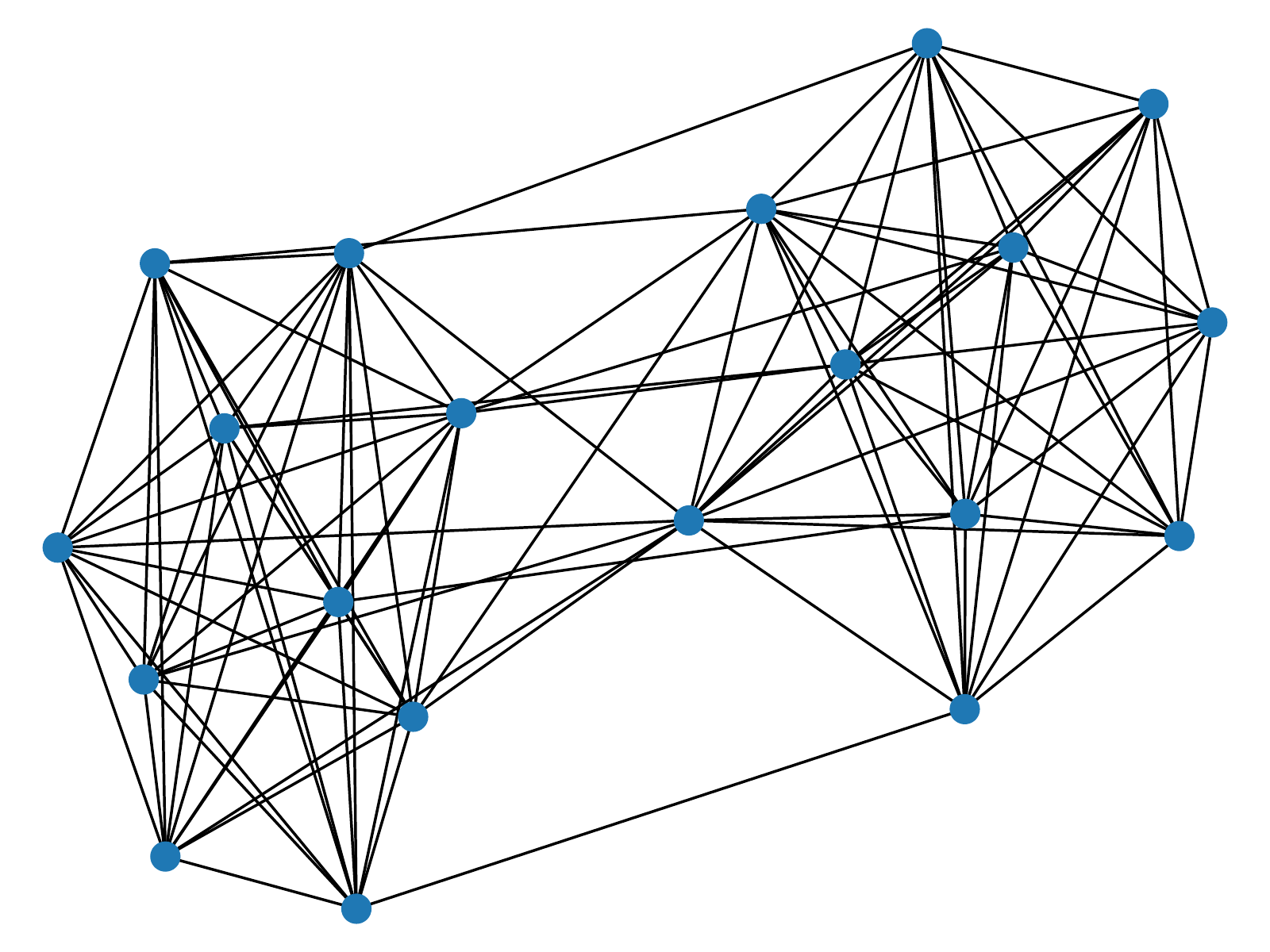}
    \caption{The undirected graph of partial nodes in HDFS data set}
    \label{ch4:fig:hdfs}
\end{figure}

Similar to the preprocessing of the LANL data set, we also remove the duplicates of the HDFS data set. Meanwhile, we integrated the events of the hosts in the $k$ step into a new security events data set. After data preprocessing, we have obtained an event data set covering 203 hosts and a total of 164 time steps. The number of events in the new data set is 51. Fig.~\ref{ch4:fig:hdfs} shows an undirected graph of some nodes in the HDFS data set.

The processed data is a sequence arranged in chronological order. Thus, we need to segment the data set for training and event prediction for the next moment. In this case, we used the sliding window to segment the data set for sequence prediction. In this way, an entire time series data can be divided into multiple sub-sequences. If the length of the sliding window is $s$, we take the first $s-1$ data as input, and the $s$-th data is used as the true value for prediction. The STEP model makes predictions based on the first $s-1$ data, and then it predicts the event at the next time step $y$. The predicted value is compared with the true value to calculate the loss. Since the predicted events in this model are all event IDs that are categorical attributes, this paper adopts cross-entropy as the loss function. The different values of the sliding window length $s$ may affect the accuracy of event prediction, which will be discussed in detail in the following experiments.

\subsection{Experimental Results}
In this section, we first define the metric to measure the performance of the model, i.e., the accuracy of event prediction. The metric accuracy is defined as follows:
\begin{equation}
Acc=\frac{\sum_{i\in N}s\left ( p_{i},y_{i} \right )}{\left | N \right |}
\end{equation}
where $p_{i}$ is the event predicted by the host $i$, $y_{i}$ is the actual event that occurred on the host $i$, and $\left | N \right |$ is the sum of events occurring on all hosts. $s\left (p_{i},y_{i} \right )$ determines whether the predicted event $p_{i}$ on the host $i$ is equal to the real event $y_{i}$. If $p_ {i}=y_{i}$, then the value of $s\left (p_{i},y_{i} \right )$ is 1, otherwise it is 0. 

In order to better evaluate the proposed model, we choose the existing two mainstream prediction models LSTM and ConvLSTM~\cite{ConvLSTM2015} as baseline methods.

\textbf{LSTM}: LSTM is a popular recurrent neural network model in the field of deep learning. It can learn long-term dependencies, especially in time series prediction tasks.

\textbf{ConvLSTM}: ConvLSTM is a recurrent neural network used for spatial-temporal prediction. It has a convolution structure in both input-to-state transition and state-to-state transition.
\begin{figure*} [htb]  
\centering
\subfigure[s=5]{
\begin{minipage}[t]{0.4\textwidth}
\includegraphics[width=1\textwidth]{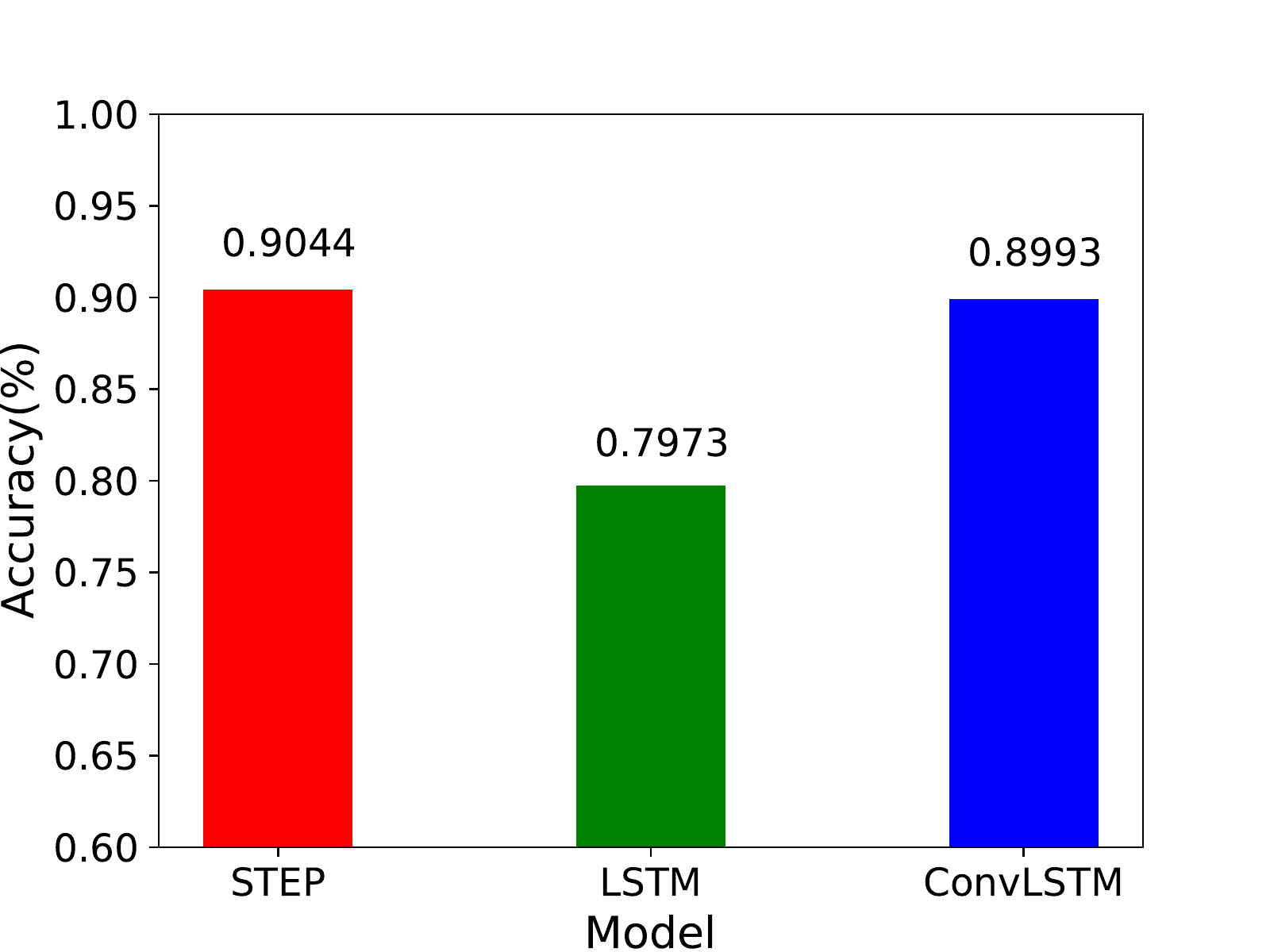} \\
\end{minipage}
}
\subfigure[s=10]{
\begin{minipage}[t]{0.4\textwidth}
\includegraphics[width=1\textwidth]{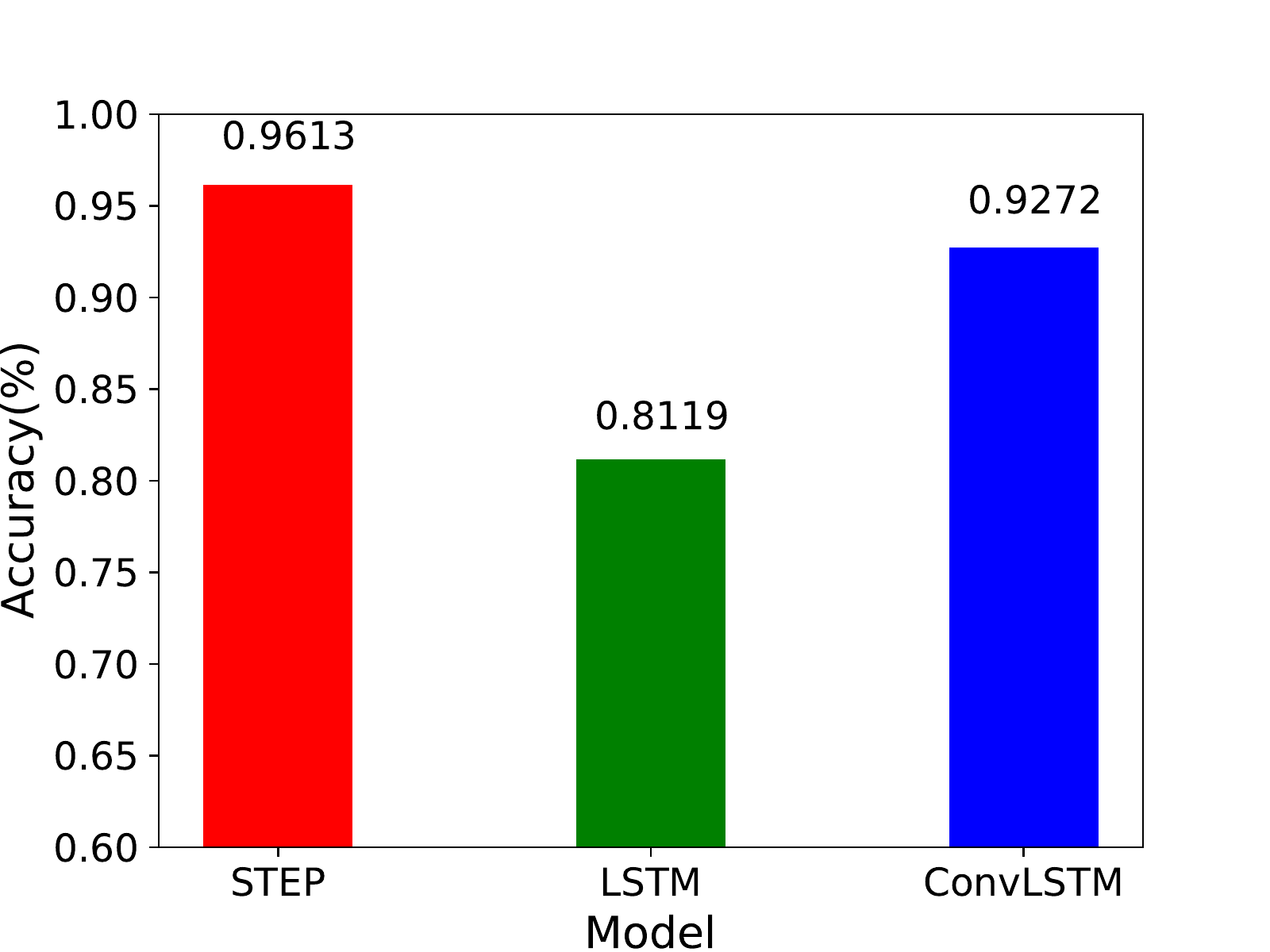} \\
\end{minipage}
}
\subfigure[s=15]{
\begin{minipage}[b]{0.4\textwidth}
\includegraphics[width=1\textwidth]{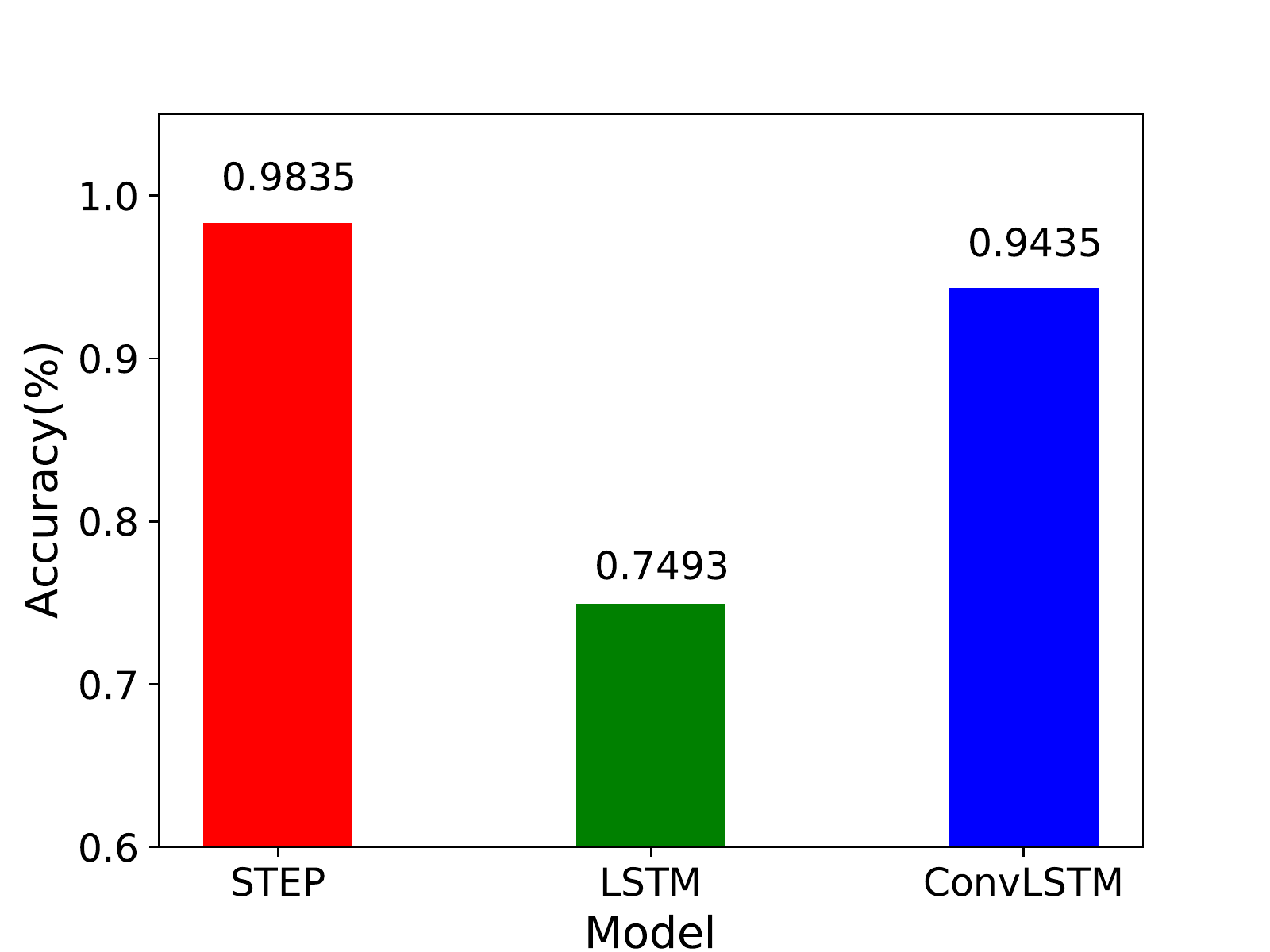} \\
\end{minipage}
}
\subfigure[s=20]{
\begin{minipage}[b]{0.4\textwidth}
\includegraphics[width=1\textwidth]{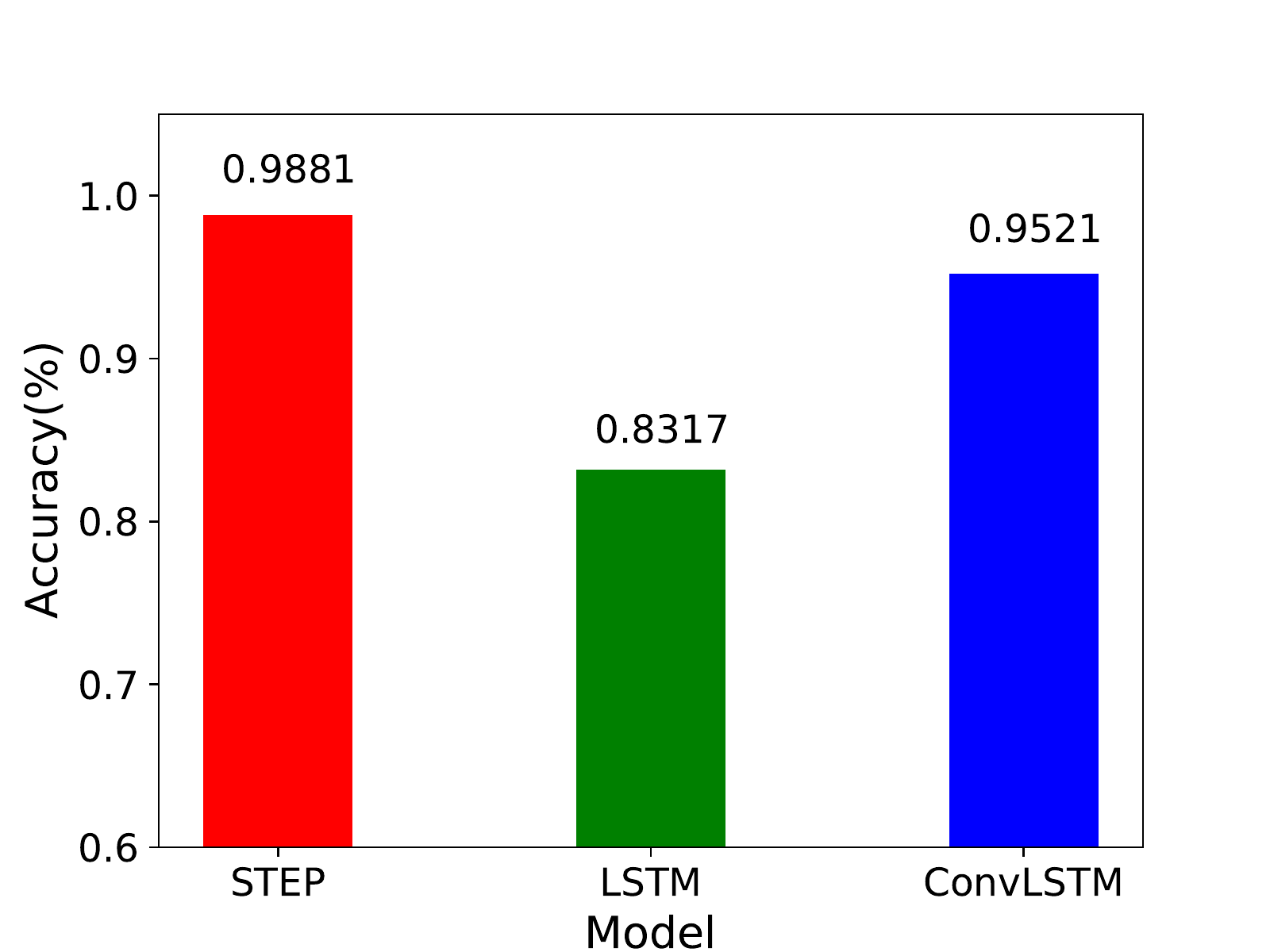} \\
\end{minipage}
}
\caption{Prediction accuracy under different sliding window sizes in HDFS data set}
\label{fig:hdfsacc}
\end{figure*}

After segmenting the original event data with a sliding window, we take 80$\%$ of them as the training set, and 20$\%$ of them as the testing set. The weight decay is set to 1.5e-3, the learning rate is 1e-3, and the dimension of the hidden layer is 128. First, we conducted experiments with the HDFS data set. The total number of input nodes is 203, the total number of events is 51, the batch size is set to 16. We run the STEP model with 100 epochs. In time series forecasting tasks, different lengths of the sliding window $s$ often affect the accuracy of the prediction results, so we discuss the influence of different sliding window sizes on the experimental results.

Specifically, we control $s$ to 5, 10, 15, 20, and record the accuracy of event prediction in these four different situations. In addition, in order to compare the prediction accuracy of different models, we also performed the same operation on the benchmark models, like LSTM and ConvLSTM. We kept the same initialization parameters, and we also set the batch size to 16. After 100 epochs, we recorded the accuracy of event prediction under different models. Figure \ref{fig:hdfsacc} records the comparisons of prediction accuracy of several models under different sliding window sizes.

As demonstrated in Fig.~\ref{fig:hdfsacc}, the prediction accuracy under the LSTM model basically remains around 80$\%$ when $s$ takes different values. In contrast, the prediction accuracy of the ConvLSTM model is higher. As the sliding window $s$ gradually increases, the prediction accuracy also shows an increasing trend, from 89.93$\%$ to 95.21$\%$. The STEP model achieve the highest prediction accuracy. Even though the sliding window is 5, it can achieves an accuracy rate of 90.44$\%$. Its prediction accuracy increases with the increase of the sliding window. If $s$ is 20, the prediction accuracy of the STEP model can reach 98.81$\%$. 

\begin{figure*}[htb]  
\centering
\subfigure[s=5]{
\begin{minipage}[t]{0.4\textwidth}
\includegraphics[width=1\textwidth]{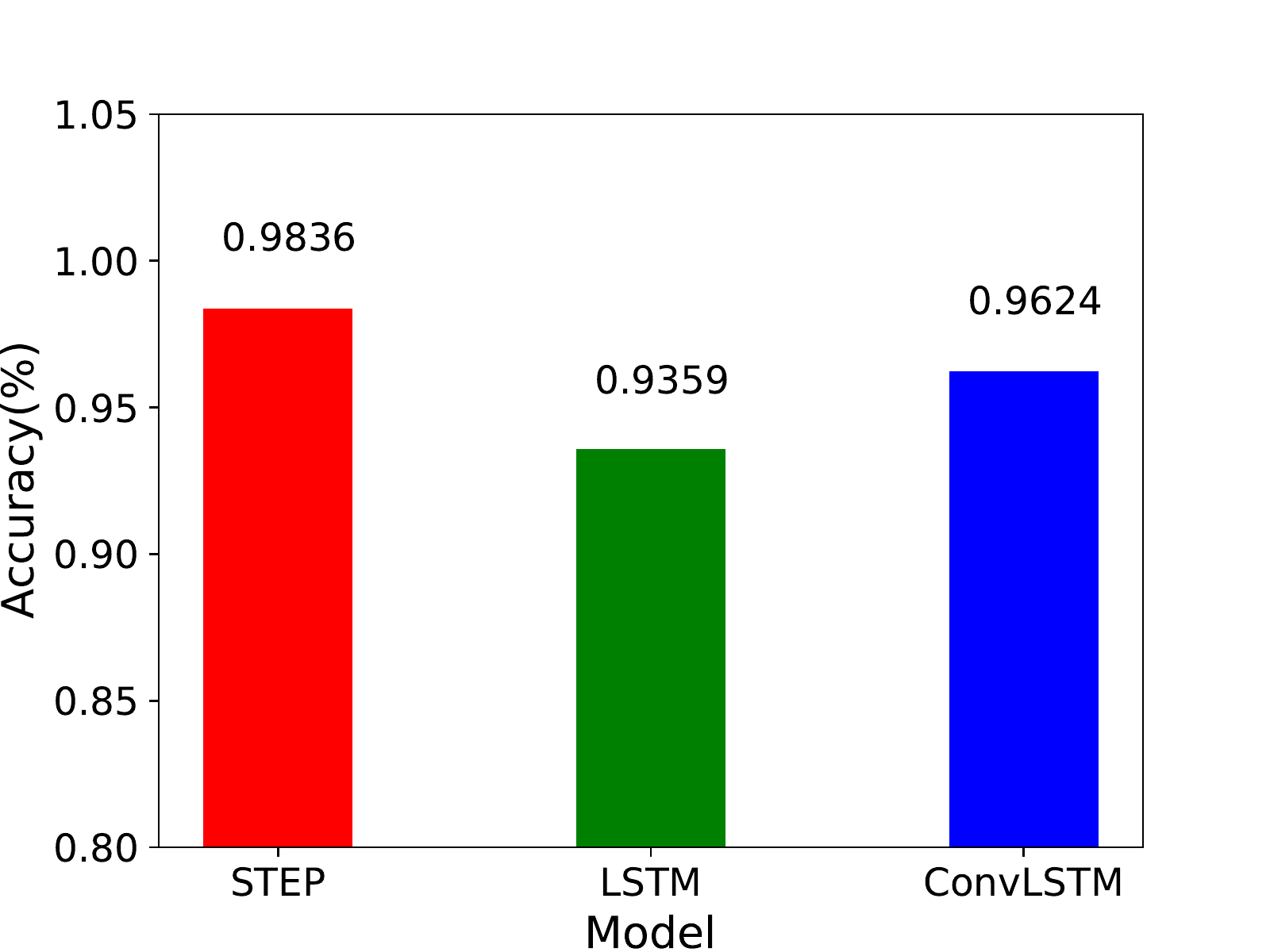} \\
\end{minipage}
}
\subfigure[s=10]{
\begin{minipage}[t]{0.4\textwidth}
\includegraphics[width=1\textwidth]{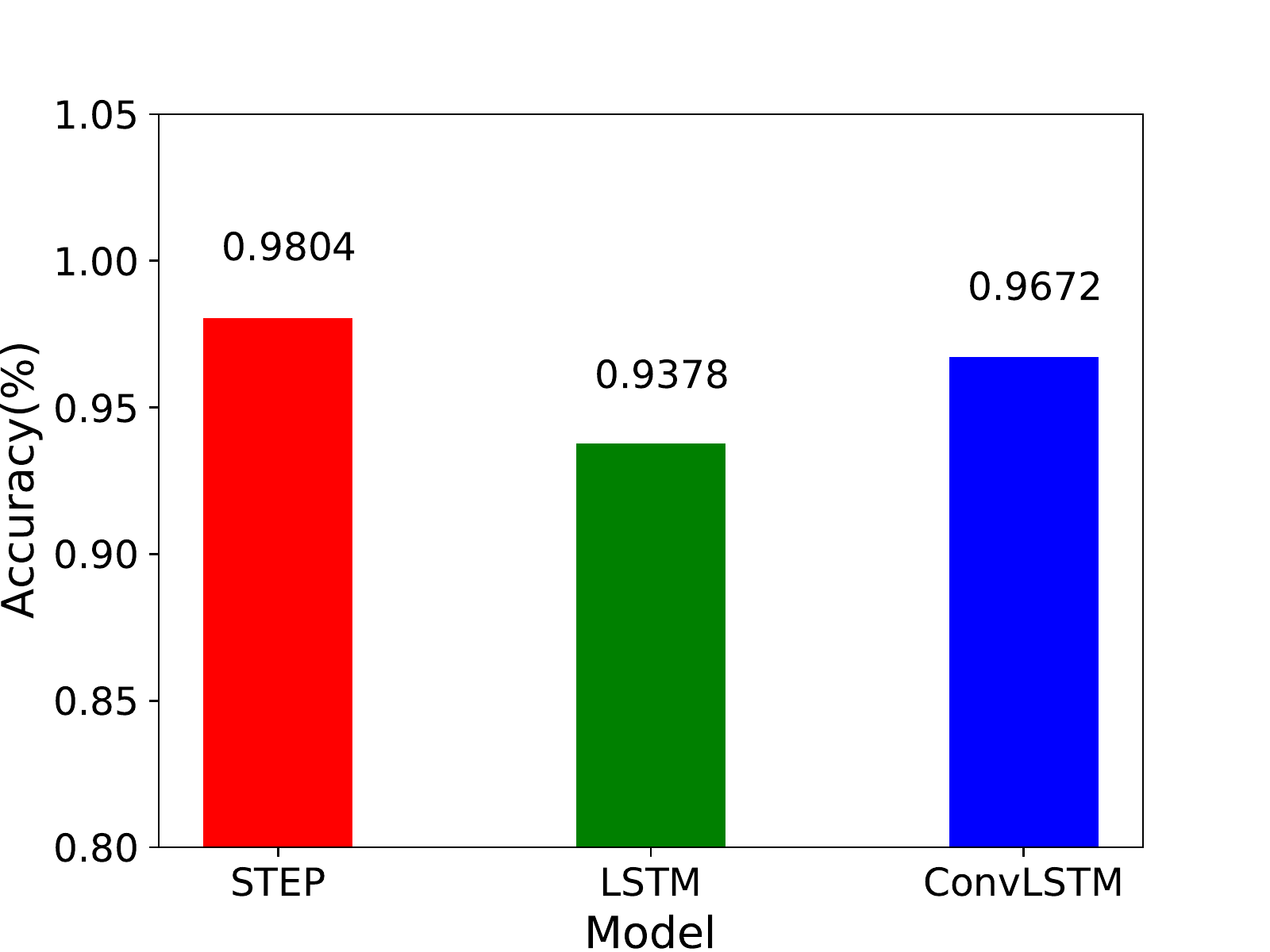} \\
\end{minipage}
}
\subfigure[s=15]{
\begin{minipage}[b]{0.4\textwidth}
\includegraphics[width=1\textwidth]{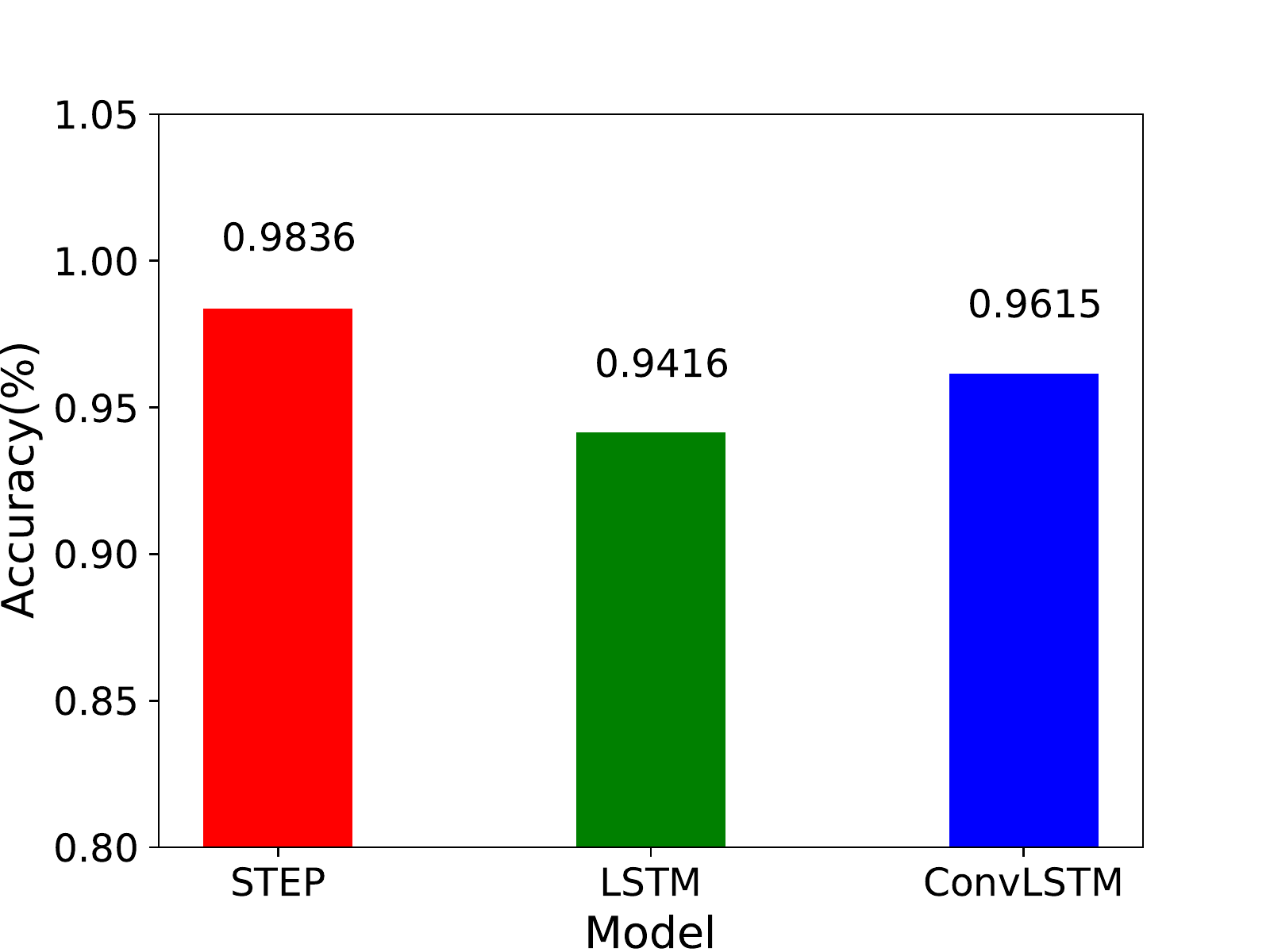} \\
\end{minipage}
}
\subfigure[s=20]{
\begin{minipage}[b]{0.4\textwidth}
\includegraphics[width=1\textwidth]{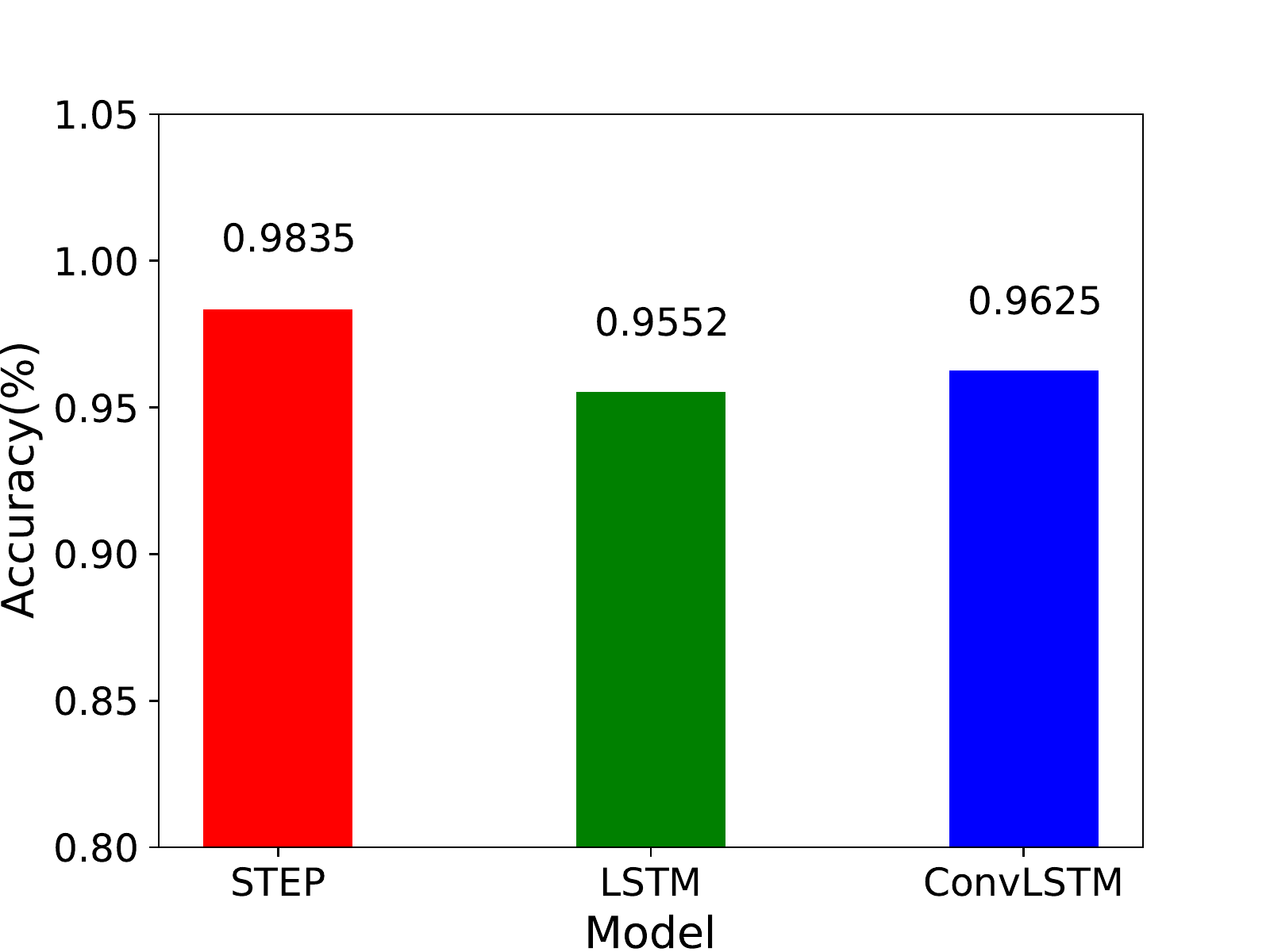} \\
\end{minipage}
}
\caption{Prediction accuracy under different sliding window sizes in LANL data set}
\label{fig:lanlacc}
\end{figure*}

Generally speaking, the accuracy of the STEP model predicting events is higher than that of ConvLSTM and LSTM, and higher accuracy can be achieved regardless of the size of the sliding window. The slightly poor prediction accuracy of LSTM lies in that it only considers the temporal characteristics of the data, and does not consider the characteristics of the spatial correlations between hosts. We have adopted an operation similar to image processing for the ConvLSTM in the experiment. We regard the feature matrix composed of different hosts at each moment as an image at that moment. However, this kind of spatial-temporal analysis does not consider the relationship between the hosts.

Similarly, we conduct the same comparative experiment under the LANL data set. We still take the size of the sliding window as 5, 10, 15, 20, and then record the prediction accuracy of events under the STEP, LSTM, and ConvLSTM models. Fig.~\ref{fig:lanlacc} shows the prediction accuracy of different models with different sizes of the sliding window. 

Similar to the results of the HDFS data set, the STEP model still has the highest accuracy, followed by the ConvLSTM model. In general, these three models have achieved an accuracy rate of more than 90$\%$ at different values of $s$. The accuracy of the STEP model is 3 to 5 percentage higher than that of LSTM and about 2$\%$ higher than that of ConvLSTM. 

The high prediction accuracy is related to the sparse events in the LANL data set. Specifically, only a small number of hosts generate events at each moment, and the remaining hosts do not generate events, which is a '0 event'. When the event ID is used to construct the feature matrix of the event, it will cause a sparse matrix.  Although we integrate the events of $k$ into a new time step to reduce the '0 event' in the feature matrix, there are still some in the new event feature matrix after data preprocessing. The overall accuracy will be higher when making specific predictions. In contrast, there are fewer '0 event' in the HDFS data set, and the accuracy of the prediction is more suitable for the actual scenario.

Although the STEP model has achieved high accuracy in predicting events, the model still has some limitations. First of all, our model is based on a static graph, that is, the relationship between hosts in the network is an undirected graph pre-specified based on historical information. In the actual network scenarios, as events change, new edges may appear in the graph. Therefore, our future work will focus on how to make correct predictions for network security incidents in the dynamic graph, so as to formulate more appropriate response measures for the network.

\section{Conclusion}
\label{conclusion}
In this paper, we propose a network security event prediction model based on temporal and spatial characteristics, STEP. Existing network security event prediction models are mostly based on time characteristics, using common LSTM or hidden Markov models for prediction. However, this method that only relies on time characteristics cannot capture the spatial dependency between network hosts. In order to capture the correlation between events, this paper proposes a spatial-temporal prediction model. STEP first uses the network history log to construct an undirected graph between network hosts, and uses the undirected graph as prior knowledge for graph convolution operations to capture the spatial characteristics between the hosts. Through graph convolution calculation, nodes in the graph can aggregate the information of neighboring nodes to form a new node representation. In addition, STEP uses graph convolution operations in LSTM to simultaneously capture the temporal characteristics of network security events. This paper utilizes the public data set to verify the proposed scheme. The experimental results show that the accuracy of predicting events under the STEP scheme is higher than that of benchmark models such as LSTM. In the future, we will focus on exploring predictions for network security events in the dynamic graph.

\bibliographystyle{IEEEtran}


\end{document}